\documentclass[journal=apchd5,manuscript=article]{achemso}

\usepackage[version=3]{mhchem} 
\usepackage{float}
\usepackage{comment}
\usepackage{esint}
\usepackage{tabularx}
\usepackage{soul}
\usepackage{amsmath}
\usepackage{amssymb}

\newcommand{\pvec}[1]{\vec{#1}\mkern2mu\vphantom{#1}}
 
\newcommand{\edit}[1]{{{\color{black}{#1}}}}  
\newcommand{\editt}[1]{{{\color{black}{#1}}}}

\usepackage[utf8]{inputenc}
\usepackage[T2A,T1]{fontenc}
\usepackage[english]{babel}

\usepackage{color}
\usepackage{ulem}
\usepackage{xcolor}
\usepackage{comment}

\usepackage{xfrac}

\usepackage{gensymb}

\usepackage[caption=false]{subfig}

\usepackage{graphicx}
\usepackage{dcolumn}
\usepackage{bm}

\author{Sergey A. Dyakov}
\affiliation{Skolkovo Institute of Science and Technology, Bolshoy Boulevard 30, str. 1, Moscow 143025, Russia}
\email{s.dyakov@skoltech.ru}
\author{Natalia Salakhova}
\affiliation{Skolkovo Institute of Science and Technology, Bolshoy Boulevard 30, str. 1, Moscow 143025, Russia}
\author{Alexey V. Ignatov}
\affiliation{Skolkovo Institute of Science and Technology, Bolshoy Boulevard 30, str. 1, Moscow 143025, Russia}
\author{Ilia M. Fradkin}
\affiliation{Skolkovo Institute of Science and Technology, Bolshoy Boulevard 30, str. 1, Moscow 143025, Russia}
\altaffiliation{Center for Photonics and 2D Materials, Moscow Institute of Physics and Technology, Institutskiy pereulok 9, Moscow Region 141701, Russia}
\author{Vitaly P. Panov}%
\affiliation{Department of Electrical and Computer Engineering, Sungkyunkwan University, Suwon, Gyeonggi-do 16419, Republic of Korea} 
\author{Jang-Kun Song}%
\affiliation{Department of Electrical and Computer Engineering, Sungkyunkwan University, Suwon, Gyeonggi-do 16419, Republic of Korea} 
\author{Nikolay A. Gippius}
\affiliation{Skolkovo Institute of Science and Technology, Bolshoy Boulevard 30, str. 1, Moscow 143025, Russia}

\title[Chiral light in twisted Fabry-Pérot cavities]
  {Chiral light in twisted Fabry-Pérot cavities}

\abbreviations{LH, RH}

\begin{document}

\begin{tocentry}

    \includegraphics[width=1\linewidth]{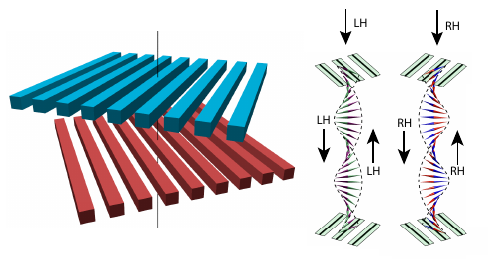}

\end{tocentry}

\begin{abstract}
Fundamental studies of the interaction of chiral light with chiral matter are important for the development of techniques that allow handedness-selective optical detection of chiral organic molecules. One approach to achieve this goal is the creation of a Fabry-Pérot cavity that supports eigenmodes with a desired electromagnetic handedness, which interacts differently with left and right molecular enantiomers. In this paper, we theoretically study chiral Fabry-Pérot cavities with mirrors comprising one-dimensional photonic crystal slabs made of van der Waals As$_2$S$_3$, a material with one of the highest known in-plane anisotropy. By utilizing the anisotropy degree of freedom provided by As$_2$S$_3$, we design Fabry-Pérot cavities with constitutional and configurational geometrical chiralities. We demonstrate that in cavities with constitutional chirality, electromagnetic modes of left or right handedness exist due to the chirality of both mirrors, often referred to as handedness preserving mirrors in the literature. At the same time, cavities with configurational chirality support modes of both handednesses due to chiral morphology of the entire structure, set by the twist angle between the optical axes of the upper and lower non-chiral anisotropic mirrors. The developed chiral Fabry-Pérot cavities can be tuned to the technologically available distance between the mirrors by properly twisting them, making such systems a prospective platform for the coupling of chiral light with chiral matter.
\end{abstract}

\section{Introduction}
Chirality is a fundamental concept that refers to the property of an object to exist in two mirror-image forms that cannot be superimposed onto each other by rotations or translations. These mirror-image forms are known as enantiomers. The scientific definition of chirality originated more than a hundred years ago \cite{kelvin1894molecular}, and it engaged increasing interest in recent decades due to the discovery of its fundamental role in nature. Chirality plays a significant role in various fields, including chemistry, biology, and physics. It influences the behavior of molecules, the properties of materials, and also the interactions between light and matter.

Chiral matter refers to materials that consist of a) chiral molecules, e.g. molecules which cannot be superimposed onto their mirror image, for example, amino acids and sugars \cite{nguyen2006chiral, Deutsche1969} or b) non-chiral molecules arranged in a chiral configuration, for example, liquid crystals \cite{Panov2003,Panov2006} \footnote{Actually, liquid crystal molecules are also chiral, but optical effects of configurational chirality can significantly dominate over constitutional chirality when the periodicity of the structure is close to the wavelength used.}. On the other hand, the simplest example of chiral light is a circularly polarized plane wave which can be either right-handed, in the case of counterclockwise rotation of the electric field vector when seen along the direction towards where the wave is propagating, or left-handed, in the case of clockwise rotation. An arbitrary electromagnetic field can be chiral too (i.e. vortex beams, interfering circularly polarized plane waves, etc.), and the quantitative measure of this a chirality density and chirality flow density \cite{bliokh2011characterizing}.  

Interaction of chiral light with chiral matter provides access to several interesting optical effects, such as circular birefringence, circular dichroism, and chiral symmetry breaking \cite{barron2009molecular, inoue2004chiral}. Circular dichroism refers to the difference of absorption of left-handed and right-handed circularly polarized light by chiral molecules, and can result in a difference in the transmitted or reflected light intensity. Circular dichroism spectroscopy is a powerful technique used to study the structure and conformational changes of chiral molecules \cite{Johnson1988, Hendry2010, Tang2011}. Circular birefringence is the ability of chiral matter to transmit left-handed and right-handed circularly polarized light with difference velocities, which results in rotation of the polarization plane of linearly polarized light as it passes through. The magnitude and the direction of optical rotation depend on the molecular structure and concentration of chiral molecules in the material \cite{inoue2004chiral, hodgkinson2001inorganic}. Finally, chiral symmetry breaking refers to a phenomenon when chiral light can induce preferential formation of one chiral form over the other. This phenomenon is of great interest in the fields of asymmetric synthesis and chiral catalysis, as it can influence the outcome of chemical reactions \cite{McConnell2007}.

The study of optical chirality is not only important for understanding the fundamental principles of light-matter interaction but also has applications in various fields, including chemistry, biology, materials science, and optics \cite{tang2010optical, genet2022chiral, baranov2023toward}. It provides a unique way to study and manipulate the properties of chiral molecules and materials, leading to advancements in drug development, molecular sensing, and various optical devices \cite{Johnson1988, Tang2011, Shapiro2003, McConnell2007,Ma2013,chambers1999stratified, Baranov2019}. When designing optical devices that utilize chirality effects, it is important to create a resonator for chiral light to enhance the interaction of light and matter. 

Resonators supporting chiral light can be constructed using resonant particles and metasurfaces. As shown in Refs.~\citenum{Tang2011, Mohammadi2019, Wu2013, Graf2019, dyakov2021photonic, yoo2015chiral}, even geometrically non-chiral structures can support a chiral electromagnetic field. In this case, however, the effect is local, meaning that the volume-average chirality density is small. Another approach to obtaining chiral light includes using spiral or helically shaped metasurfaces, which introduce a twist into the system \cite{Kan2015, FernandezCorbaton2019, Hendry2010, Pham2016, Wang2023, Konishi2011, Barbillon2020}. Although such structures can routinely support chiral eigenmodes, the spatial localization of chiral light in them is still confined to a two-dimensional surface. With that in mind, three-dimensional Fabry-Pérot cavities formed by a pair of mirrors are of particular interest. It has been demonstrated that Fabry-Pérot cavities can provide local predominance of one handedness over another \cite{feis2020helicity, Sun2022, gautier2022planar, mauro2023chiral} or even support a standing wave of a particular handedness \cite{Voronin2022}. The second case is especially interesting because it allows obtaining null-free resonant modes with a high chirality density in the entire gap between the mirrors. Regrettably, Fabry-Pérot resonators supporting a field of a certain handedness cannot be formed merely by a pair of homogeneous isotropic plates, as the latter do not preserve the handedness of light upon reflection.

Refs.~\citenum{li2020spin, Semnani2020, gorkunov2021bound, Plum2015, Voronin2022, Liu2020} provide a solution to this problem by demonstrating so-called handedness-preserving mirrors. One common method to achieve handedness-preserving mirrors is through the use of quasi-2D chiral materials or metasurfaces. These objects, having low symmetry, possess intrinsic chirality and thus interact differently with left-handed and right-handed circularly polarized light, resulting in the preservation of the polarization state upon reflection. Made from chiral constituents, these resonators exhibit constitutional chirality. While two-dimensionally periodic handedness-preserving mirrors are undoubtedly strong candidates for a platform for chiral molecular photonics \cite{baranov2023toward} and can be used to engineer single-handed optical cavities \cite{Voronin2022}, an obvious drawback of these mirrors is the complexity of fabrication and positioning within a Fabry-Pérot resonator. In this paper, we present two alternatives to two-dimensionally periodic handedness-preserving mirrors.

The first alternative is based on the engineering of one-dimensionally periodic handedness-preserving mirrors made of homogeneous stripes of anisotropic material. The symmetry of the geometrically non-chiral shape of this structure can be reduced by tilting the optical axis of the anisotropic material away from parallel with the direction of the stripes.

The second alternative to two-dimensionally periodic handedness-preserving mirrors involves a paradigm shift in how these mirrors are used. Instead of creating a chiral Fabry-Pérot cavity with chiral mirrors arranged in a non-chiral shape, our aim is to develop a chiral Fabry-Pérot cavity using non-chiral mirrors arranged in a chiral shape. In this context, we employ non-chiral, one-dimensionally periodic, horizontally oriented photonic crystal slabs that are twisted around the vertical axis. This geometrical construction can be described as configurational chirality, in contrast to constitutional chirality represented by a Fabry-Pérot resonator with chiral mirrors.

As both alternatives involve using one-dimensional photonic crystal slabs, in this paper, we present them as two parts of the most general configuration characterized by an arbitrary twist angle between two gratings and the arbitrary orientation of the optical axis of the anisotropic material (Fig.\,\ref{fig:sample}a). For both types of mirrors, we employ a recently synthesized van der Waals material, As$_2$S$_3$ \cite{slavich2023exploring}, known for having one of the highest in-plane anisotropies among natural materials. In As$_2$S$_3$ flakes, the principal axes of anisotropy are mutually orthogonal, with two of them lying in the $xy$-plane.

The paper is organized as follows. Initially, we will formulate a theory describing the eigenmodes of a Fabry-Pérot cavity with constitutional or configurational chirality. Then, we will present a specific realization of cavities of both types. Finally, we will simulate the interaction of chiral Fabry-Pérot eigenmodes with chiral matter by calculating the emissivity of chiral point light sources.
\section{Results}
\subsection{Theory}
\label{theory}
To describe eigenmodes of the Fabry-Pérot resonator formed by two mirrors of subwavelength periodicity, we use an {approach} \cite{menzel2010advanced} where the local field is expressed in terms of a complex vector {of amplitudes} written in the Cartesian basis {(denoted as "$xy$")} or circular polarization basis {(denoted as "$\rho\sigma$") as}
\begin{equation}
    \pvec{\mathrm{A}}_{xy} = 
    \begin{bmatrix}
        a_x\\a_y
    \end{bmatrix},~~~~~~~
    \pvec{\mathrm{A}}_{\rho\sigma} = 
    \begin{bmatrix}
        a_\rho\\a_\sigma
    \end{bmatrix},
\end{equation}
with a transformation matrix $\mathbb{T}$:
\begin{equation}
    \mathbb{T}\equiv\mathbb{T}_{\rho\sigma \leftarrow xy} = 
    \frac{1}{\sqrt{2}}
    \begin{bmatrix}
        1 & i \\ 1 & -i
    \end{bmatrix},
\end{equation}
such that $\pvec{\mathrm{A}}_{\rho\sigma} = \mathbb{T}\pvec{\mathrm{A}}_{xy}$. {As soon as the vector of amplitudes can describe waves propagating in either positive or negative $z$-directions, we introduce notation $\pvec{\mathrm{A}}^+$ for a positively propagating wave, $\pvec{\mathrm{A}}^-$ for a negatively propagating wave. It should be noted that the vectors $\pvec{\mathrm{A}}^+_{\rho\sigma} = [1, 0]^T$ and $\pvec{\mathrm{A}}^-_{\rho\sigma} = [0, 1]^T$ describe right-handed wave, and the vectors $\pvec{\mathrm{A}}^-_{\rho\sigma} = [1, 0]^T$ and $\pvec{\mathrm{A}}^+_{\rho\sigma} = [0, 1]^T$ describe left-handed wave.}


Eigenmodes of the Fabry-Pérot resonator can be found by solving the following eigenvalue problem:
\begin{equation}
    \label{eigenproblem}
    \mathbb{M}\pvec{\mathrm{A}}^+\equiv \mathbb{P}\mathbb{R}^\mathrm{upper}\mathbb{P}\mathbb{R}^\mathrm{lower}\pvec{\mathrm{A}}^+ = m\pvec{\mathrm{A}}^+,
\end{equation}
and setting eigenvalue $m$ to 1. In Eqn.~\ref{eigenproblem}, $\mathbb{R}^\mathrm{upper}$ and $\mathbb{R}^\mathrm{lower}$ are reflection matrices of the upper and lower mirrors and $\mathbb{P}$ is a propagation matrix:
\begin{equation}
    \mathbb{P} = 
    \begin{bmatrix}
        e^{ikg} & 0\\ 0 & e^{ikg}
    \end{bmatrix},
\end{equation}
where $k = 2\pi/\lambda$ is an absolute value of the wave vector. {The sequence} of matrices in Eqn.~\eqref{eigenproblem} suggests that the eigenvector $\pvec{\mathrm{A}}^+$ corresponds to wave taken in the proximity of the lower mirror. To describe the electromagnetic field of the eigenmode {in vacuum}, {we calculate the following parameters:
\begin{align}
    \label{chirden2}
    \begin{split}
    I = &~ (\vec{E}\cdot\vec{E}^*)+(\vec{H}\cdot\vec{H}^*)\\
    C = &~ \mathrm{Im}(\pvec{E} \cdot \pvec{H}^*),
    \end{split}
\end{align}
where the scalar product of complex vectors $\vec{a}$ and $\vec{b}$ is defined as $(\vec{a}\cdot\vec{b})\equiv\sum_i a_ib_i$. The parameter $I$ has a meaning of the field intensity, while $C$ refers to the chirality density \cite{bliokh2011characterizing, fernandez2016objects}.}

Using expressions for electric and magnetic field vectors in terms of the components of the vector $\pvec{\mathrm{A}}$,
\begin{align}
    \pvec{E} = 
    \begin{bmatrix}
        a_x^+\\a_y^+\\0    
    \end{bmatrix}, ~~~~~
    \pvec{H} = 
    \begin{bmatrix}
        -a_y^+\\a_x^+\\0    
    \end{bmatrix}, ~~~~~  & \text{for~} k_z>0\\
    \pvec{E} = 
    \begin{bmatrix}
        a_x^-\\a_y^-\\0    
    \end{bmatrix}, ~~~~~
    \pvec{H} = 
    \begin{bmatrix}
        a_y^-\\-a_x^-\\0    
    \end{bmatrix}, ~~~~~  & \text{for~} k_z<0,
\end{align}
one can express the time average normalized intensity and chirality density for the superposition of two waves propagating in opposite directions in the gap between two mirrors as
\begin{align}
    \label{chirden}
    \begin{split}
    I = &~ |a_\rho^-|^2 + |a_\rho^+|^2 + |a_\sigma^-|^2 + |a_\sigma^+|^2\\
    C = &~ |a_\rho^-|^2 - |a_\rho^+|^2 - |a_\sigma^-|^2 + |a_\sigma^+|^2.
    \end{split}
\end{align}
where signs "+" and "-" correspond to waves propagating in positive and negative $z$-directions, respectively.

\begin{figure*}
    \centering
    \includegraphics[width=1\linewidth]{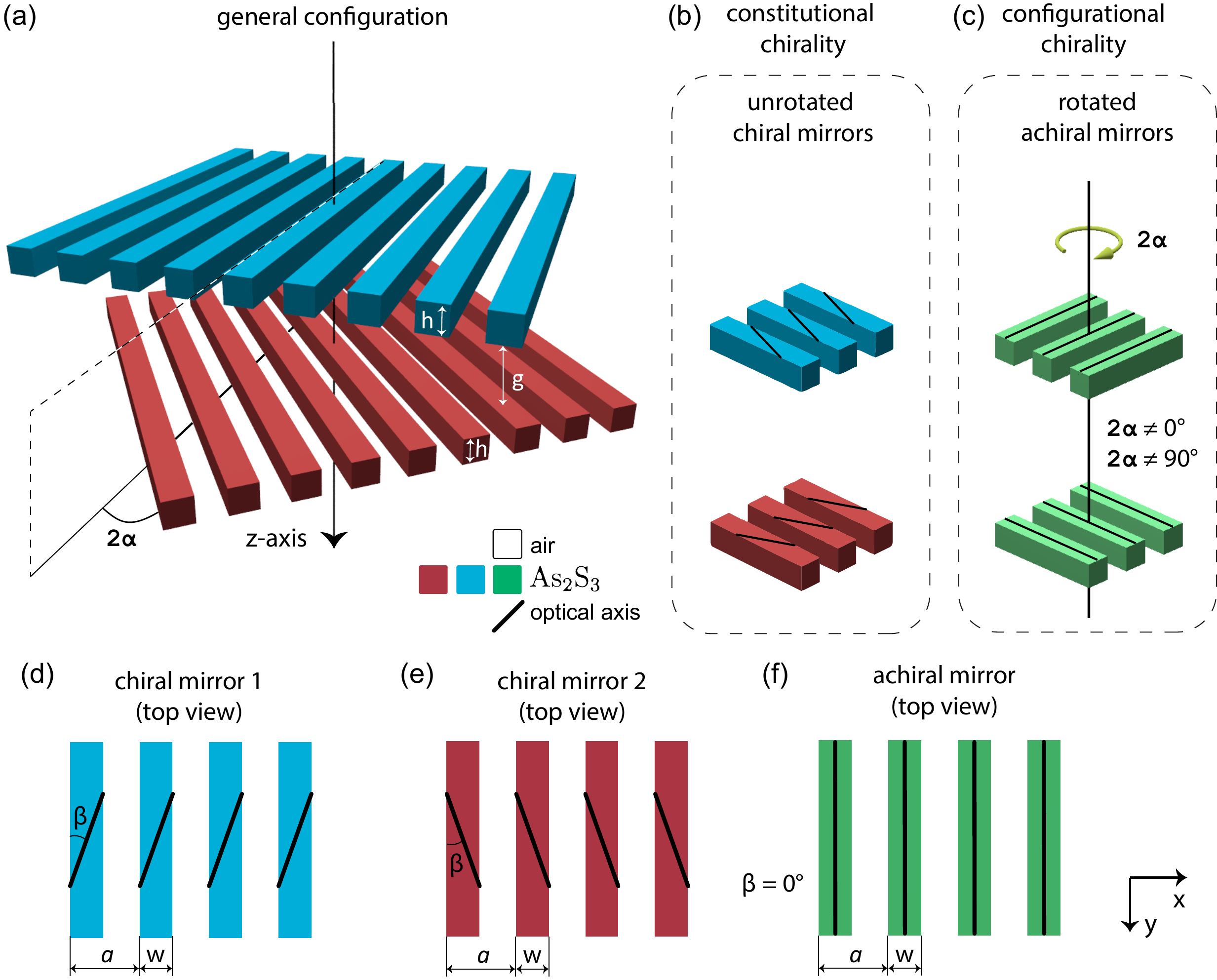}
    \caption{(a) The most general configuration of a chiral resonator with twisted one-dimensional gratings. (b) Chiral resonators formed by chiral and non-chiral mirrors. (c) A top view of the unrotated chiral and non-chiral mirrors. In (a)--(f) color represents the orientation of optical axis.}
    \label{fig:sample}
\end{figure*}

Let us now consider a system of two chiral mirrors constituting a Fabry-Pérot resonator supporting a particular handedness. Similar to {Refs.~\citenum{Semnani2020, Voronin2022}}, {we choose the upper and lower mirrors in such a way that} the reflection matrices of the upper and lower mirrors will have the following form:
\begin{equation}
    \label{matr12}
    \mathbb{R}^\mathrm{upper}_{\rho\sigma} = 
    \begin{bmatrix}
        0 & 0\\ r & 0
    \end{bmatrix}, ~~~~~~
    \mathbb{R}^\mathrm{lower}_{\rho\sigma} = 
    \begin{bmatrix}
        0 & r\\ 0 & 0
    \end{bmatrix},    
\end{equation}
with $|r|=1$. Although such a Fabry-Pérot resonator already supports eigenmodes of a particular handedness {\cite{Semnani2020, Voronin2022}}, for better control of chiral light-matter interaction, we introduce an additional degree of freedom by rotating the upper and lower mirrors by angles $\alpha$ and $-\alpha$ ($0\le\alpha\le \pi/2$) about the $z$-axis. The upper and lower reflection matrices in such a system will be
\begin{equation}
    \widetilde{\mathbb{R}}_{\rho\sigma}^\mathrm{upper} = \mathbb{S}\mathbb{R}^\mathrm{upper}_{\rho\sigma}\mathbb{S}^{-1}, ~~~~~ \widetilde{\mathbb{R}}_{\rho\sigma}^\mathrm{lower} = \mathbb{S}^{-1}\mathbb{R}^\mathrm{lower}_{\rho\sigma}\mathbb{S},
\end{equation}
where $\mathbb{S}$ is a rotation matrix:
\begin{equation}
    \mathbb{S}=
    \begin{bmatrix}
        \cos \alpha & \sin \alpha \\ -\sin \alpha & \cos \alpha 
    \end{bmatrix}.
\end{equation}
{In this case,} the expression for the total matrix $\mathbb{M}$ is reduced to
\begin{equation}
{\mathbb{M}}_{\rho\sigma}= \mathbb{P}\widetilde{\mathbb{R}}^\mathrm{upper}_{\rho\sigma}\mathbb{P}\widetilde{\mathbb{R}}_{\rho\sigma}^\mathrm{lower} = 
    \begin{bmatrix}
        0 & 0\\0 & r^2e^{2ikg}e^{-4i\alpha}
    \end{bmatrix},  
\end{equation}
The only non-trivial solution of the corresponding eigenproblem is the vector $\pvec{\mathrm{A}}_{\rho\sigma}=[0, 1]^T$ representing the LH wave, and the eigenvalue $m=r^2e^{2ikg}e^{-4i\alpha}$ which is equal to 1 at
\begin{equation}
\label{gk}
    g = \frac{1}{k}\left(-\mathrm{arg}~r + 2\alpha + \pi N \right),  \text{~where~}  N = 1,2,3,\ldots
\end{equation}
In accordance with formula~\eqref{chirden}, the chirality density of this eigenmode equals to $C=2$ and does not depend on the twist angle $2\alpha$. Note that the position of non-zero elements on the secondary diagonal of the upper and lower matrices \eqref{matr12} sets the handedness of the eigenmode.

\begin{figure*}
    \centering
    \includegraphics[width=1\linewidth]{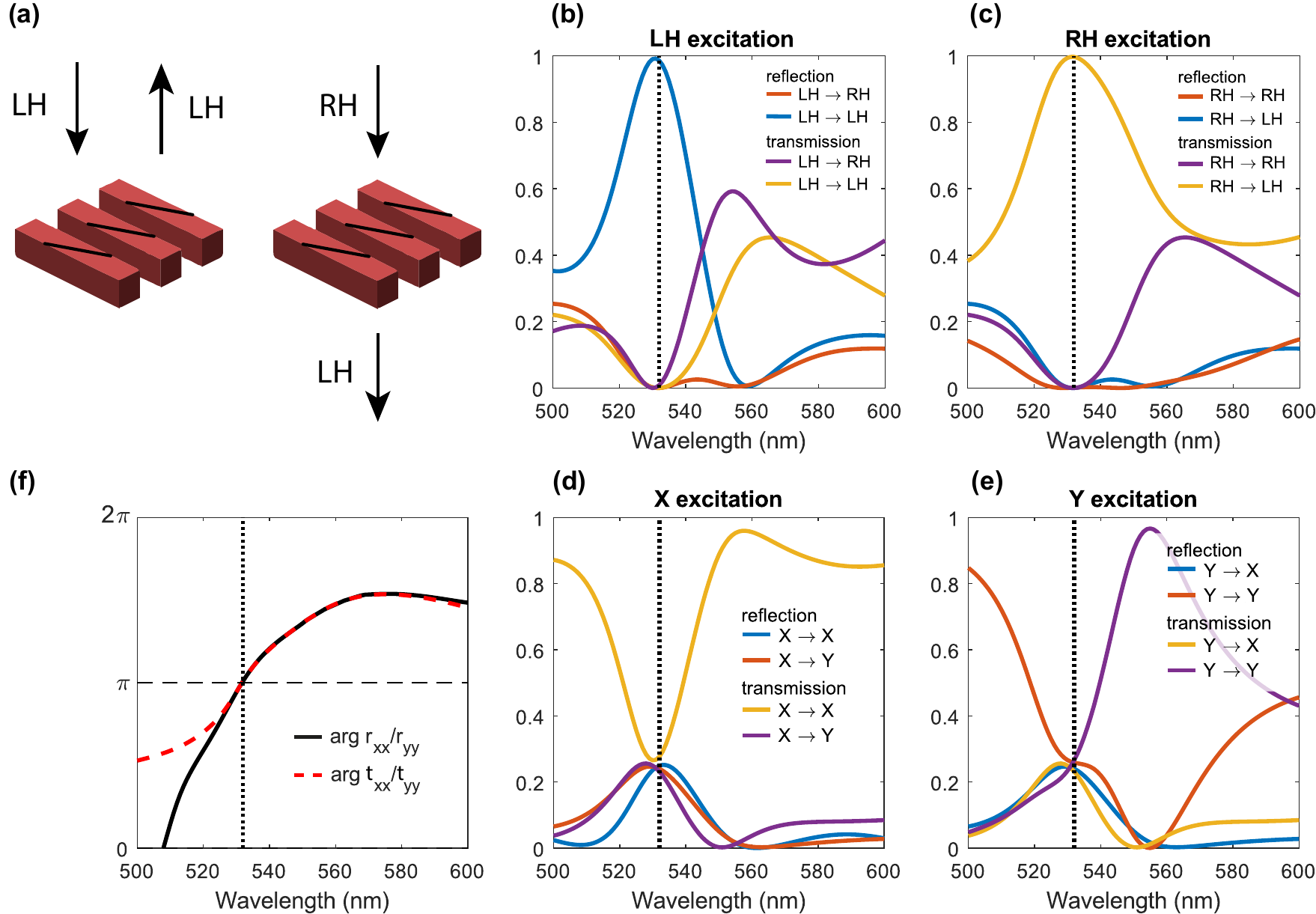}
    \caption{(a) Schematic view of the lower chiral grating, designed to reflect the LH light and transmit the RH light. (b) and (c) The cross-polarized reflection and transmission coefficients of the chiral grating for RH and LH incident light. (d) and (e) The cross-polarized reflection and transmission coefficients of the chiral grating upon X- and Y-polarized incident light. (f) The phase difference between the $r_{xx}$ and $r_{yy}$ coefficients (black solid line) and between the $t_{xx}$ and $t_{yy}$ coefficients (red dashed line).}
    \label{fig11}
\end{figure*}
\begin{figure*}[th]
    \centering
    \includegraphics[width=1\linewidth]{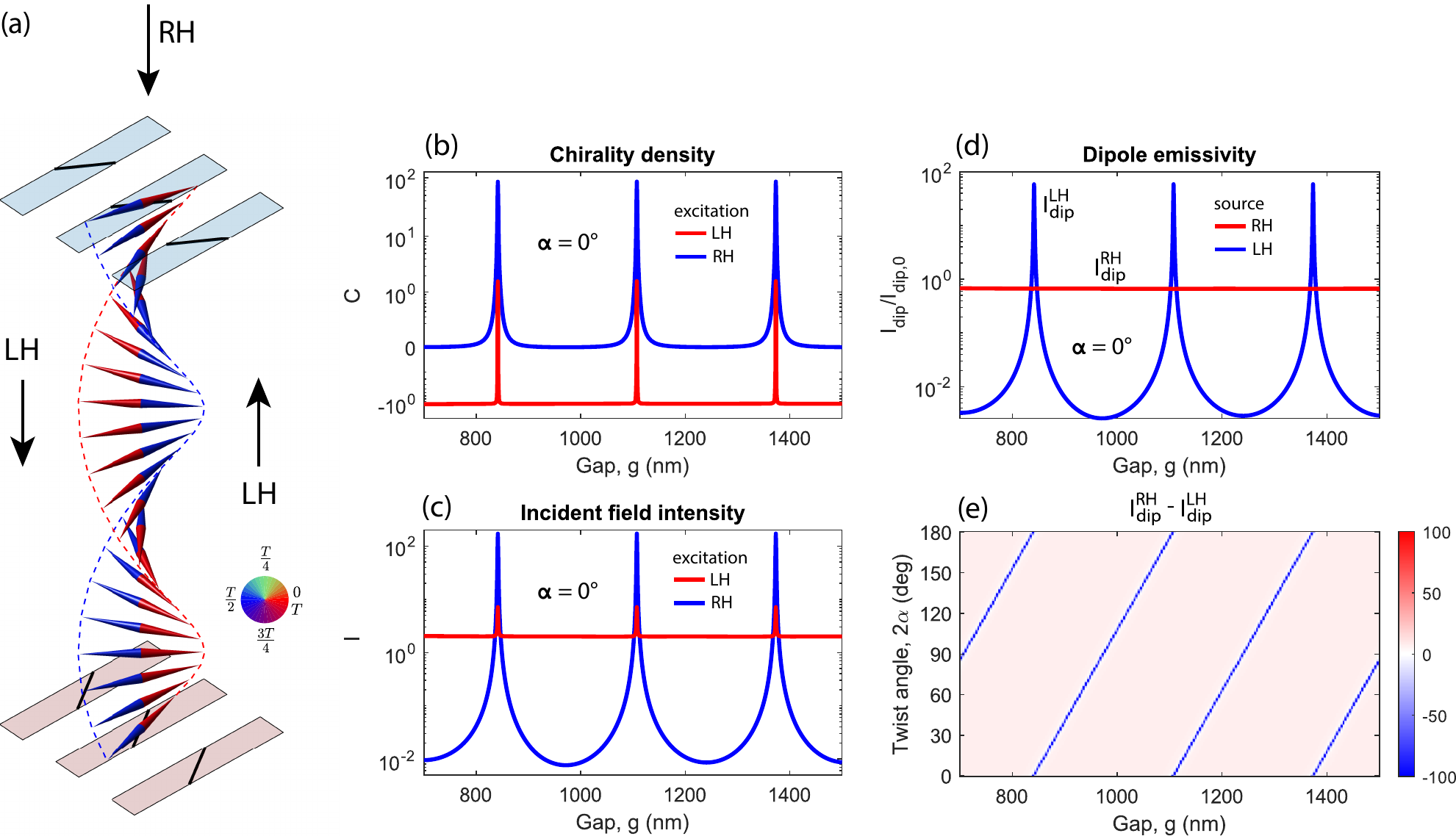}
    \caption{(a) Sketch view of a chiral resonator supporting LH Fabry-Pérot modes. {The cones represent electric vectors of the incident wave settled in the gap between the mirrors with phases specified by the circular colorbar. The field distribution is calculated for the gap size $g\approx575$\,nm and the twist angle $2\alpha=0^\circ$. The gap-size dependence of (b) the time-average normalized chirality density and (c) incident field intensity in the middle of the cavity.  (d) The gap-size dependence of the emissivity of RH and LH dipole sources placed in the middle of the cavity. (e) The gap-size and twist-angle dependence of a difference between RH and LH dipoles emissivities.}}
    \label{fig12}
\end{figure*}

Next, we consider a system of two non-chiral mirrors. Such a Fabry-Pérot resonator can have eigenmodes of a particular handedness if the mirrors are designed to reflect two orthogonal linear polarizations with a phase difference of $\pi$ \cite{Plum2015}. This situation is possible when mirrors are anisotropic in the $xy$-plane. The reflection matrix of each mirror will have the form: 
\begin{equation}
    \mathbb{R}_{xy}^0 =  
    \begin{bmatrix} 
    r & 0 \\ 0 & -r
    \end{bmatrix},
\end{equation}
that corresponds to 
\begin{equation}
    \label{form15}
    \mathbb{R}_{\rho\sigma}^0 =  
    \begin{bmatrix} 
    0 & r \\ r & 0
    \end{bmatrix},
\end{equation}
i.e. both circular polarizations are reflected contrary to the chiral mirror case. For simplicity, as in the case of chiral mirrors, we assume that $|r|=1$. To introduce chirality into the structure, we rotate the non-chiral anisotropic mirrors about the $z$-axis by angles $\alpha$ and $-\alpha$ ($0\le\alpha\le \pi/2$), and the upper and lower reflection matrices in such a system will be
\begin{equation}
    \widetilde{\mathbb{R}}_{xy}^\mathrm{upper} = \mathbb{S}\mathbb{R}^0_{xy}\mathbb{S}^{-1}, ~~~~~  \widetilde{\mathbb{R}}_{xy}^\mathrm{lower} = \mathbb{S}^{-1}\mathbb{R}^0_{xy}\mathbb{S}.
\end{equation}
After some algebra, we obtain the following expression for the matrix $\mathbb{M}$ in circular basis:
\begin{equation}
    \mathbb{M}_{\rho\sigma} = 
    \begin{bmatrix}
        r^2e^{2ikg}e^{+4i\alpha} & 0\\ 0& r^2e^{2ikg}e^{-4i\alpha}
    \end{bmatrix}.
\end{equation}
The corresponding eigenvalue problem has two series of non-trivial solutions
\begin{align}
    \pvec{\mathrm{A}}_{\rho\sigma}^{(1)} &= \begin{bmatrix} 0\\1 \end{bmatrix},~~~~~~ m^{(1)}=r^2e^{2ikg}e^{+4i\alpha}\\
    \pvec{\mathrm{A}}_{\rho\sigma}^{(2)} &= \begin{bmatrix} 1\\0 \end{bmatrix},~~~~~~ m^{(2)}=r^2e^{2ikg}e^{-4i\alpha},
\end{align}
which are degenerate at $\alpha=0$, $\pi/4$ and $\pi/2$. When the degeneracy is lifted, the system is chiral, and at gap sizes
\begin{equation}
    \label{finalg}
    g = \frac{1}{k}\left(-\mathrm{arg} ~r \mp 2\alpha + \pi N\right)
\end{equation}
it supports either left or right handedness ($C=\pm 2$). In formula~\eqref{finalg}, $N$ is a positive integer, and signs "+" and "-" correspond to LH and RH modes respectively.

\subsection{Cavity with constitutional chirality}
\label{constchirsec}
We start with a cavity of constitutional chirality, i.e. with the cavity formed by chiral mirrors. We are particularly interested in one-dimensional photonic crystal slab mirrors because they are easier to fabricate \edit{as} compared to two-dimensional metasurfaces. Let us consider a mirror that consists of an array of infinite stripes aligned in the $y$-direction and periodically arranged along the $x$-direction, as shown in Fig.\,\ref{fig:sample}b,d,e. A one-dimensional array of isotropic stripes with a uniform profile along the $y$-direction has $D_{2h}$ rotational symmetry and therefore is non-chiral. \edit{Using the anisotropy degree of freedom provided by As$_2$S$_3$, we reduce the rotational symmetry of one-dimensionally periodic mirrors from D$_{2h}$ to C$_{2h}$ which leads to a planar chirality of the structure. In this configuration, the in-plane optical axis is oriented at an angle $\beta$ with the direction of stripe, where $\beta$ must not be equal $0$ or $\pm \pi/2$.} For the proof-of-principle demonstration of the existence of eigenmodes with a specific handedness in this structure, we choose a target wavelength $\lambda=532$~nm as it corresponds to the spectral maximum of the in-plane anisotropy of As$_2$S$_3$. The dielectric tensor components of As$_2$S$_3$ at $\lambda=532$~nm are as follows: $\varepsilon_{xx} = 3.23^2$, $\varepsilon_{yy} = 2.91^2$, $\varepsilon_{zz} = 2.50^2$. (See Supporting Information for the spectra of the dielectric permittivity tensor components.)  

To satisfy requirement \eqref{matr12} for the reflection matrices of the upper and lower mirrors, the grating should be designed in such a way that normally incident circularly polarized light reflects with the same handedness, while the transmitted light exhibits the opposite handedness, as shown in Fig.,\ref{fig11}a. Although the handedness of the cavity's eigenmode is independent of the absolute value of the amplitude reflection coefficient $r$ of the mirror, we strive to make it as close to 1 as possible in order to achieve high quality factor of the resulting Fabry-Pérot resonances. In numerical optimizations, we vary the grating period $a$, stripe thickness $h$, stripe width $w$, and the optical axis orientation angle $\beta$. The optimal configuration is achieved with $a = 360$~nm, $w=0.41a$, $h = 320$~nm, and $\beta=33^\circ$. As proven in Ref.~\citenum{Voronin2022}, a necessary prerequisite for chiral photonic crystal slabs to exhibit the properties of a handedness-preserving mirror is to establish an isotropic background with equal reflectance and transmittance coefficients. This should be combined with a pair of eigenstates having identical eigenfrequencies, opposite parity, and a phase difference of $\pi/4$.

\begin{figure*}[th]
    \centering
    \includegraphics[width=1\linewidth]{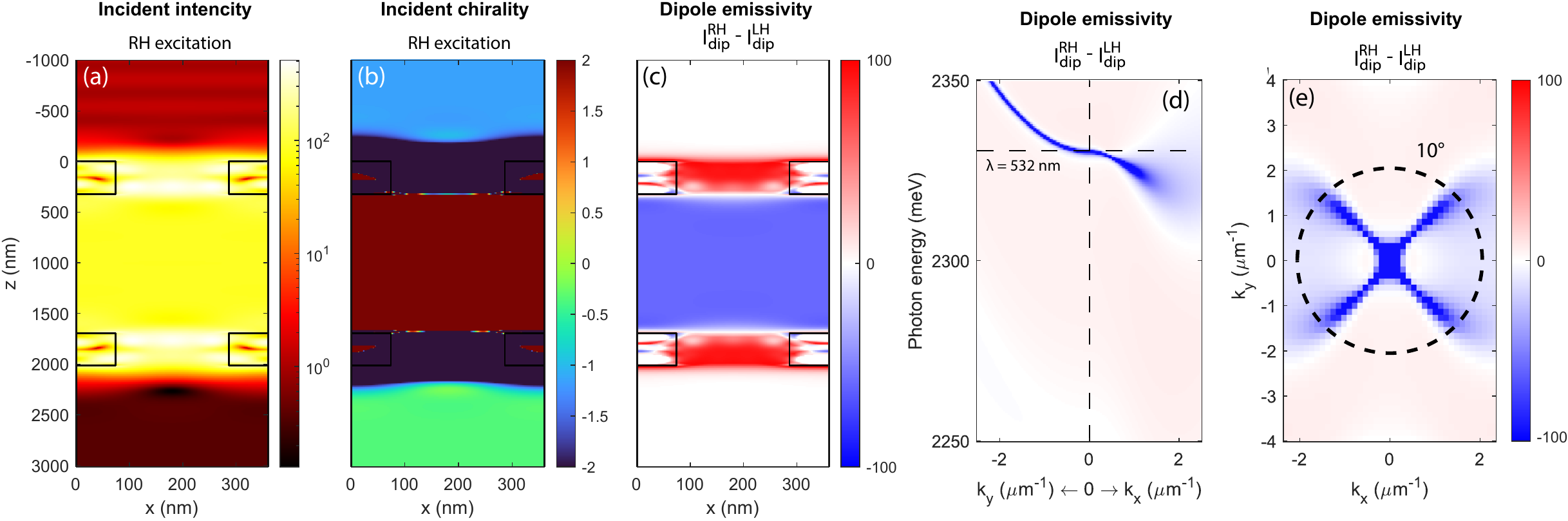}
    \caption{The spatial distribution of (a) the time-average normalized incident field intensity and (b) chirality density in the resonator with chiral mirrors supporting LH modes, calculated for RH excitation. Additionally, in (c) the spatial distribution of the difference between RH and LH dipole emissivities is shown. Panels (d) and (e) depict the wavelength and in-plane wavevector dependencies of the difference between RH and LH dipole emissivities. Calculations are made for $2\alpha=30^\circ$ and $g=1570$~nm.}
    \label{fig13}
\end{figure*}
The cross-polarization reflection and transmission spectra of the optimized mirror under RH and LH incidence are presented in Fig.\,\ref{fig11}b,c. At a wavelength of $\lambda=532$~nm, the mirror exhibits a near-perfect reflection of the normally incident LH light, while the RH incident light is almost completely transmitted. \edit{In the case of a linearly-polarized incident light, the reflection and transmission cofficients are both sufficiently non-zero.} It is worth noting that the phase difference between the cross-polarization coefficients $r_{xx}$ and $r_{yy}$, as well as between $t_{xx}$ and $t_{yy}$, equals $\pi$ (Fig.\,\ref{fig11}f). \editt{This indicates that in the optimized mirror, the light's handedness is conserved upon reflection, whereas it changes to the opposite handedness upon transmission.} 

The resulting chiral resonator, consisting of two enantiomeric forms of the optimized chiral mirror, should support the LH mode, which can be excited from the far field by the RH wave. \edit{As shown in Fig.\,\ref{fig12}a, the incident RH wave refracts into the gap region as LH wave, which subsequntly reflects from the lower mirror as LH wave, which finally reflects from the upper mirros again as LH wave. As a result of the interference between LH waves propagating in the positive and negative $z$-directions, a standing wave is formed. As can be seen from Fig.\,\ref{fig12}a, locally this standing wave is linearly polarized, however, the polarization direction forms a helix with the $z$-coordinate. Dispite being locally linearly polarized, such a standing wave is obviously cannot be superimposed on its mirror image and, thus, is chiral.}

\edit{To quantitatively characterize the chiral properties of the resonator's eigenmodes, we expose it to a circularly polarized plane wave with unit amplitudes of electric and magnetic fields and calculate the time-averaged intensity and chirality density in the middle of the cavity\footnote{Field intensity and chirality density are expressed in Gaussian units as \cite{vazquez2018optical}:
\begin{equation}
\label{expI}
    I = \frac{1}{16\pi}\left(\pvec{E}\cdot\pvec{E}^* + \pvec{H}\cdot\pvec{H}^*\right)
\end{equation}
\begin{equation}
\label{expC}
    C = \frac{\omega}{2c^2}~\mathrm{Im}\left(\pvec{E}\cdot\pvec{H}^*\right).
\end{equation}
In plotting $I$ and $C$ we omit the factors $1/16\pi$ and $\omega/2c^2$ and set the incoming amplitudes of electric and magnetic fields to 1.} using formula~\eqref{chirden}. From Fig.\,\ref{fig12}b,c it is evident that the RH excitation induces a positive chirality density of the electromagnetic field inside the gap. \edit{This observation, along with the convention for LH and RH light accepted in this paper,} indicates that the RH incidence wave indeed excites the LH mode. Due to the almost perfect reflection of the \edit{LH} wave, the incident intensity and chirality density reach values of the order $10^2$ at resonant gap sizes defined by formula \eqref{gk} with $\alpha=0$. \edit{Because specific values of the parameters $I$ and $C$ are defined by the precision at which Eqn.~\eqref{matr12} is satisfied, further geometrical optimization may lead to even higher intensity and chirality density in the gap.}}

\editt{It should be noted that although the coefficients $t_\mathrm{RH\rightarrow LH}$ and $t_\mathrm{LH\rightarrow LH}$ are both close to 0 in the optimized upper chiral mirror, the ratio $|t_\mathrm{RH\rightarrow LH}|^2/|t_\mathrm{LH\rightarrow LH}|^2$ appeared to be as high as $\approx 100$. Due to this, the LH mode is much more effectively excited by the RH wave, than by the LH wave. The corresponding peak values of the field intensity and chirality density in Fig.\,\ref{fig12}b,c are almost 2 orders of magnitude lower compared to RH excitation.}

Further, the chiral electromagnetic field supported by the Fabry-Pérot resonator in question should interact differently with left\edit{-handed} and right\edit{-handed} chiral matter. 
\edit{In the most general case of chiral and anisotropic media, the constitutive relations of the electromagnetic field encompass macroscopic coefficients of chirality and bianisotropy \cite{lindell1994electromagnetic}, which must be taken into account when solving Maxwell's equations. In this study, we focus on matter with constitutional chirality and assume that the concentration of chiral molecules is so low that the macroscopic chirality and bianisotropy coefficients are negligible.}

\begin{figure*}
    \centering
    \includegraphics[width=1\linewidth]{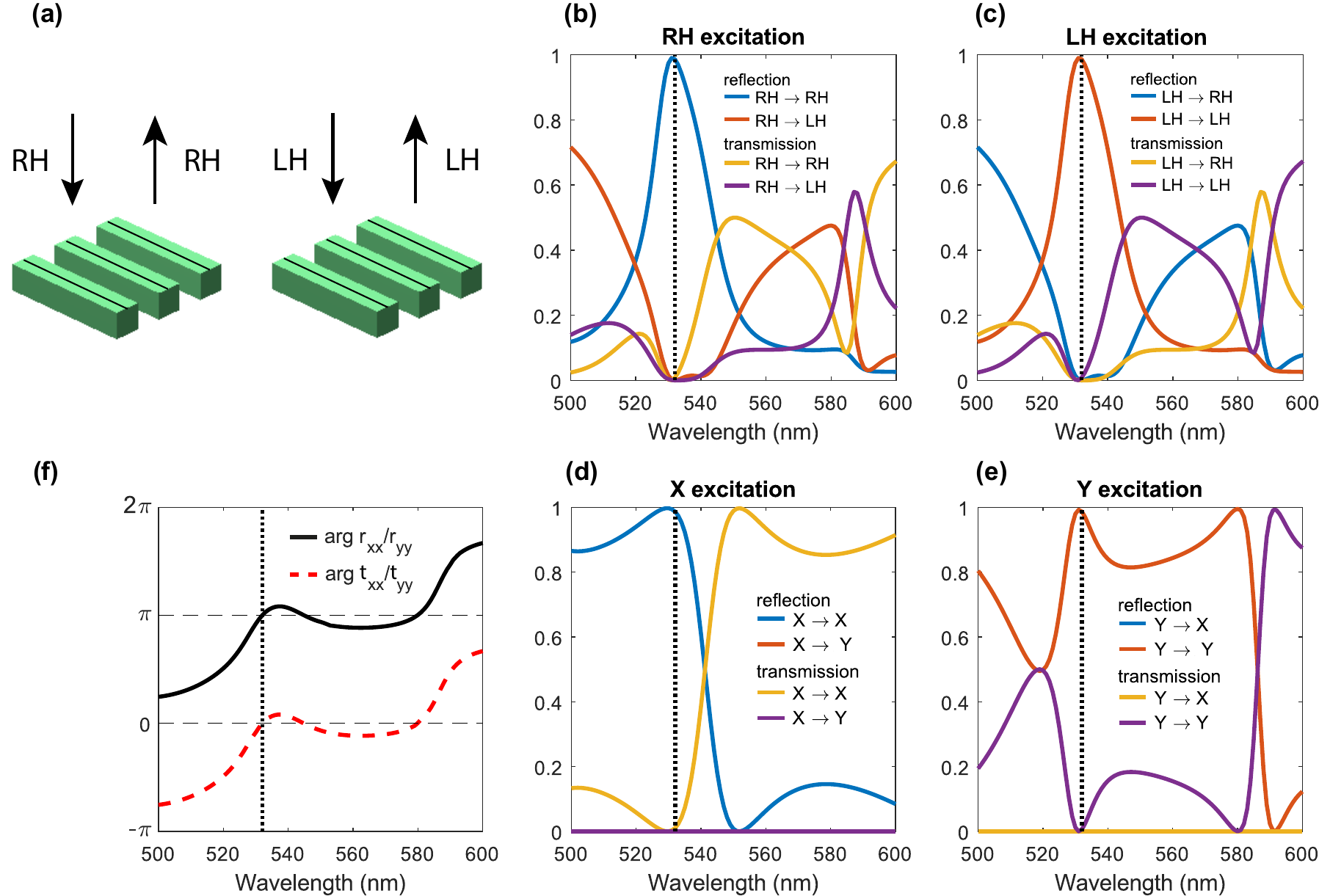}
    \caption{(a) Schematic view of a non-chiral grating designed to reflect either right handedness or left handedness. (b) and (c): The cross-polarized reflection and transmission coefficients of the chiral grating calculated for RH and LH incident light. (d) and (e): The cross-polarized reflection and transmission coefficients of the chiral gratingcalculated for X- and Y-polarized incident light. (f) The phase difference between the $r_{xx}$ and $r_{yy}$ coefficients (represented by the black solid line) and between the $t_{xx}$ and $t_{yy}$ coefficients (represented by the red dashed line).}
    \label{fig21}
\end{figure*}
To simulate the interaction of a chiral electromagnetic field, supported by an optimized Fabry-Pérot resonator, with chiral matter, we introduce a chiral emitter into the system, placing it in the middle of the gap between the mirrors. The chiral emitter is modeled by radiation of oscillating electric and magnetic point dipole moments with a phase difference of $\pm\pi/2$ \cite{Voronin2022, schäfer2023chiral}:
\begin{equation}
\label{pmdipoles}
\pvec{p}=\pvec{p}_0e^{i\omega t}, ~~~~~~~~~~~~~ \pvec{m}=\pm ic\pvec{p}_0e^{i\omega t},
\end{equation}
where $c$ is the speed of light and $\pvec{p}_0$ is the amplitude of electric dipole moment oscillations. In \eqref{pmdipoles}, the signs "+" and "-" correspond to right-handed and left-handed emitters, respectively. The details of calculation of electric and magnetic dipoles' radiation in terms of the Fourier modal method are presented in Supporting Information. In the following, we will calculate the far-field emissivity of the chiral emitters, defined as the radiation intensity of the emitter in the cavity normalized to the radiation intensity of the same emitter in the vacuum.

The emissivity of LH and RH emitters \edit{averaged over the $x$-, and $y$-orientations of the vector $\pvec{p}_0,$} placed in the middle of the gap between the mirrors with optimized geometry is shown in Fig.\,\ref{fig12}d. One can see that at resonant gap sizes, the LH emitter couples effectively to LH eigenmodes of the resonator, and the normalized emissivity reaches $\approx 100$. In contrast, the emissivity of the RH emitter is almost constant and equals $\approx 0.8$, indicating that the RH emitter is almost unaffected by the resonator supporting LH modes.

Formula \eqref{gk} suggests that the resonant gap size can be tuned by rotating one chiral mirror relative to the other one. To demonstrate this fact in our system, we calculate the emissivity of LH and RH emitters placed in the resonator in which the upper and lower chiral gratings are rotated about the $z$-axis by angles $+\alpha$ and $-\alpha$ (relative to their initial orientation along the $y$ axis), respectively. Fig.\,\ref{fig12}e shows that, indeed, the resonant gap size depends linearly on the twist angle $2\alpha$, in accordance with the electric vector orientation at different $z$-coordinates in the chiral standing wave as shown in Fig.\,\ref{fig12}a. \editt{The possibility to adjust the resonant gap size by a rotation of one grating relative to another is important from a practical viewpoint, because it makes resonant conditions technologically more available.}

\begin{figure*}[th]
    \centering
    \includegraphics[width=1\linewidth]{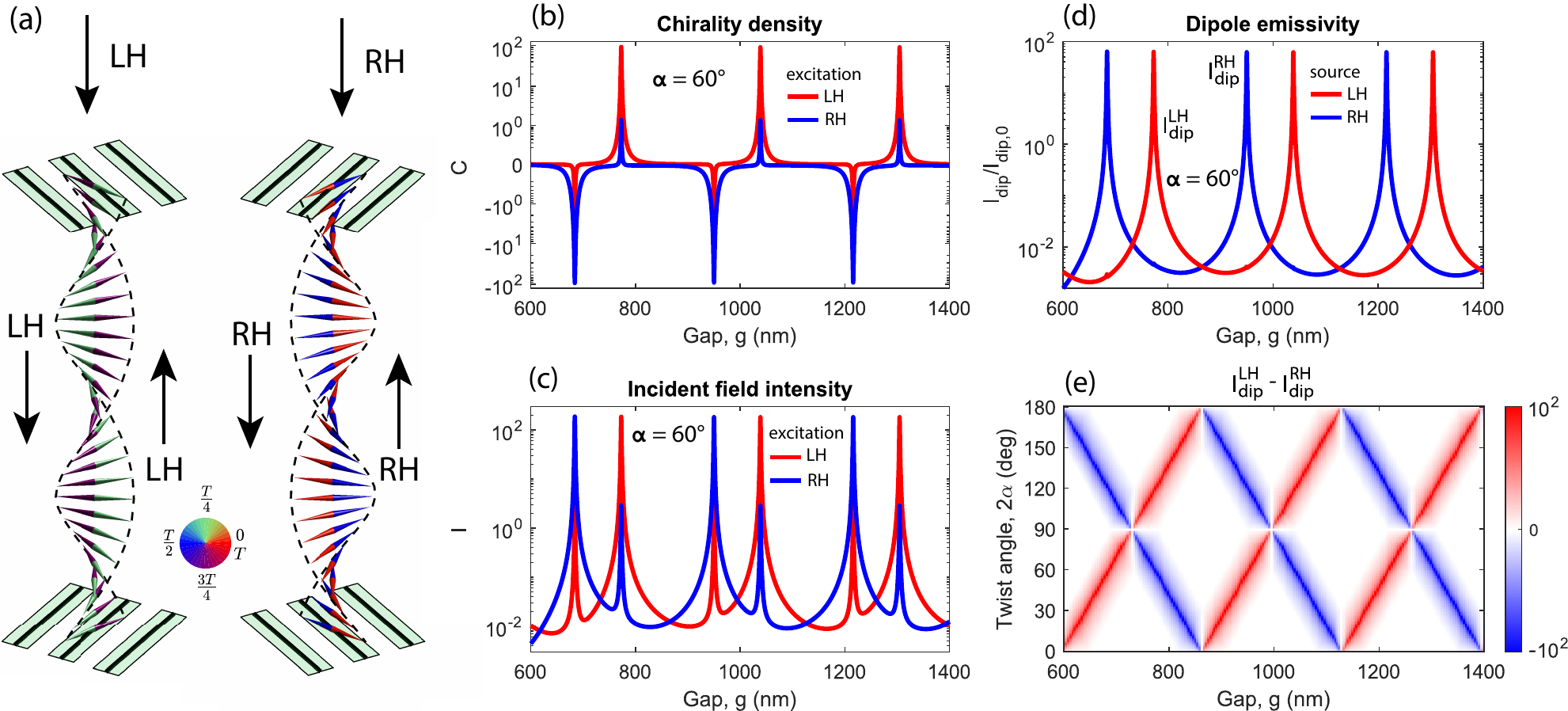}
    \caption{(a) The sketch view of a resonator with non-chiral mirrors supporting Fabry-Pérot modes with right handedness or left handedness. \edit{The cones represent electric vectors of the incident wave settled in the gap between the mirrors, with phases specified by the circular colorbar. The field distributions are calculated for the twist angles $2\alpha=120^\circ$ (left figure) and $2\alpha=60^\circ$ (right figure), and the gap size $g\approx684$\,nm. The gap-size dependence of (b) the time-average normalized chirality density  and (c) incident field intensity calculated for RH and LH incident waves. (d) The gap-size dependence of the emissivity of RH and LH dipole sources placed in the middle of the cavity. (e) The gap-size and twist-angle dependence of a difference between RH and LH dipoles emissivities.}}
    \label{fig22}
\end{figure*}
From a more practical prospective, it is important that the incident field intensity $I$, the chirality denisty $C$, and the dipole's emissivity $I_\mathrm{dip}$ shown in Fig.~\ref{fig12} for the middle of the cavity, exhibit spatial uniformity. Fortunately, according to formula \eqref{chirden}, in a standing wave formed by two counter-propagating circularly-polarizing waves of the same handedness, there is no $z$-dependence of the parameters $I$, $C$. Therefore, no dependence on the $z$-coordinate is expected for the emissivity as well. Our full wave calculations reveal that in the designed resonator, there is no spatial modulation of the incident field intensity and chirality density not only in the $z$-direction but also in all three directions (see Fig.\,\ref{fig13}a,b). In the vicinity of the mirrors, however, the parameters $I$ and $C$ differ from those in the middle of the cavity due to the influence of the mirrors' near-field. A similar pattern is observed for the dependence of the far-field emissivity on the emitter's position: the LH emitter radiates 100 times more power than the RH emitter almost everywhere in the gap, except for the regions near the mirrors (see Fig.\,\ref{fig13}c).

\edit{Finally, the dispersion of the far-field emissivity near the $\Gamma$-point of the photonic crystal lattice is shown in Fig.\,\ref{fig13}d,e. One can see that for the resonator to maintain the desired properties, the wavelength must not deviate too much from the target value, and the emission angle must not be too far from normal.}


\subsection{Cavity with configurational chirality}
Next, we consider a Fabry-Pérot cavity with a configuration chirality, i.e., formed by non-chiral mirrors twisted in a chiral configuration, as shown in Fig.\,\ref{fig:sample}c. To maintain consistency throughout this paper, we have selected  As$_2$S$_3$ chalcogenide as the material for the non-chiral mirrors, like for chiral ones. However, in this case, to achieve a non-chiral morphology of the photonic crystal slab, the As$_2$S$_3$ optical axis is aligned parallel to the stripes (Fig.\,\ref{fig:sample}f). It's important to note that, as outlined in Sec.~\ref{theory}, the only essential condition for non-chiral mirrors to preserve handedness upon reflection is that the phase difference between the amplitude reflection coefficients $r_{xx}$ and $r_{yy}$ is $\pi$. Therefore, there is no necessity to use an anisotropic material as the stripes material. In this regard, the results presented below can be replicated on a broader range of platforms, such as silicon-on-insulator wafers or simply homogeneous layers of anisotropic material.

\begin{figure*}[th]
    \centering
    \includegraphics[width=1\linewidth]{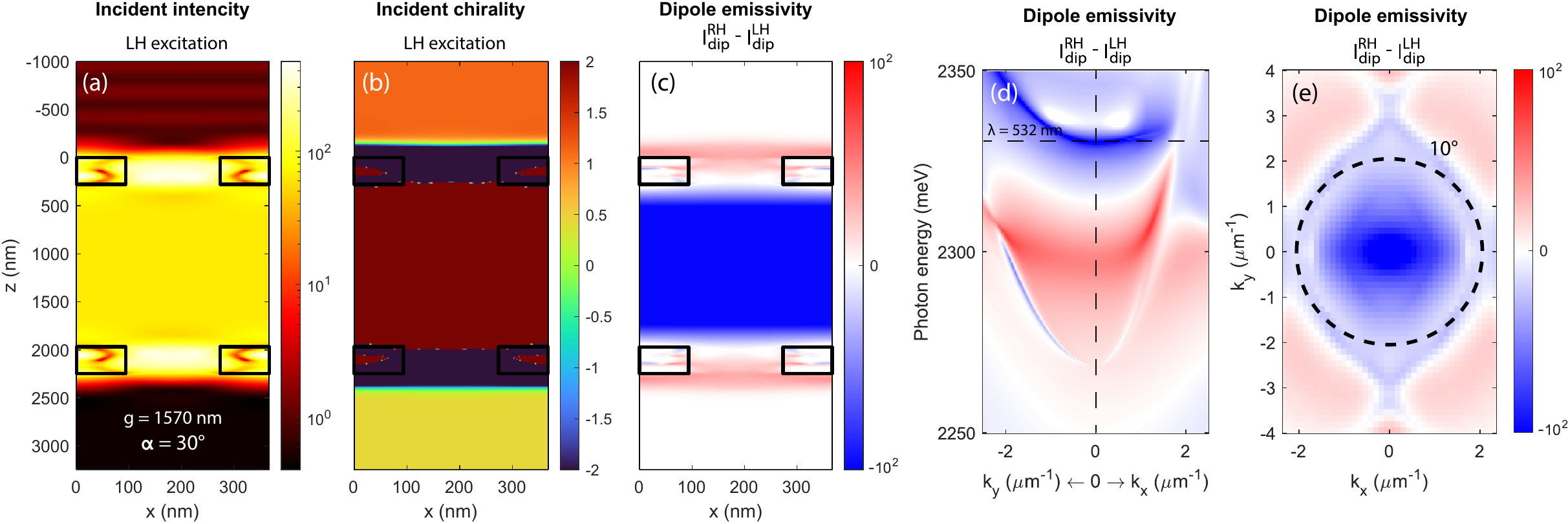}
    \caption{The spatial distribution of (a) the time-average normalized incident field intensity and (b) chirality density in the resonator with non-chiral mirrors supporting LH modes, calculated for LH excitation. Additionally, in (c) the spatial distribution of the difference between RH and LH dipole emissivities is shown. Panels (d) and (e) depict the wavelength and in-plane wavevector dependencies of the difference between RH and LH dipole emissivities. Calculations are made for $2\alpha=0$ and $g=1373$~nm.}
    \label{fig23}
\end{figure*}

In optimizing the structural geometry, in addition to satisfying the phase difference condition of the reflection coefficients, we also require that the absolute values $|r_{xx}|$ and $|r_{yy}|$ be close to 1 to obtain high-quality Fabry-Pérot modes. The variation of the grating period $a$, stripe width $w$, as well as the stripe thickness $h$, leads to the following optimal configuration: $a=353$~nm, $w=0.55a$, $h=388$~nm. As depicted in Fig.~\ref{fig21}f and Fig.~\ref{fig21}d,e, with these parameters, the phase difference between the amplitude reflection coefficients $r_{xx}$ and $r_{yy}$ is indeed close to $\pi$, and their absolute values are close to 1. On the other hand, the phase difference between the amplitude transmission coefficients appears to be close to 0. These observations indicate that in the optimized non-chiral mirrors, the handedness of light is preserved both in reflection and transmission. Furthermore, as shown in Fig.\,\ref{fig21}b,c, in accordance with formula \eqref{form15}, in the case of optimized non-chiral mirrors, an incident wave of \textit{any} handedness is reflected with the same handedness almost perfectly, in contrast to the case of chiral mirrors. The lack of handedness-selective reflection of non-chiral mirrors under normal incidence is inherently expected and is attributed to the D$_{2h}$ symmetry of these mirrors. 

Similar to the case of chiral mirrors, the ability of non-chiral mirrors to reflect a circularly polarized wave without changing its handedness enables the existence of chiral Fabry-Pérot modes. In Fig.~\ref{fig22}a, it can be observed that in a cavity with configurational chirality, both LH and RH incident waves excite a chiral linearly polarized standing wave, which is formed by two counter-propagating circularly-polarized waves of the same handedness. The fields excited by the LH and RH waves represent two enantiomers of the same mirror-image form.

The resulting dependence of chirality density and incident field intensity on the gap size in the Fabry-Pérot cavity with configurational chirality is illustrated in Fig.~\ref{fig22}b,c. It is evident that the chirality density within the resonator varies between approximately -100 and 100, depending on the gap size. This variation reaffirms that the cavity supports both LH and RH eigenmodes, as discussed in Sec.~\ref{theory}. The resonant gap sizes for LH and RH modes are described by formula \eqref{finalg}. As depicted in Fig.~\ref{fig22}b,c, an eigenmode of a particular handedness is excited nearly 100 times more effectively by an incident wave of the same handedness than by an incident wave of the opposite handedness. This asymmetry is attributed to the earlier assertion that light transmission through the optimized non-chiral mirrors occurs without changing its handedness. More specifically, this is explained by the fact that while all the transmission coefficients are close to zero, the ratio $|t_{\mathrm{RH}\rightarrow \mathrm{RH}}|^2/|t_{\mathrm{RH}\rightarrow \mathrm{LH}}|^2$ (and its counterpart for LH excitation) is as high as approximately 80 in the optimized structure.

Like in the case of a cavity with constitutional chirality, the emissivity of a chiral emitter of a certain handedness is maximal at gap sizes corresponding to eigenmodes of the same handedness (see Fig.\,\ref{fig22}d). The resulting emissivity of \edit{both} LH and RH emitters reaches $\sim$100, but the peak emissivities are achieved at gap sizes, different for LH and RH modes, as predicted by formula \eqref{finalg}. 

To examine the impact of the twist angle between the gratings on the resonant properties of the cavity with configurational chirality, we calculate the emissivity of RH and LH emitters versus the twist angle and the gap size (Fig.\,\ref{fig22}e). At $2\alpha=0^\circ$ and $\pi/2$, i.e. when the upper and lower gratings are either parallel or perpendicular to each other, the emissivities of RH and LH dipoles are equal (white regions at the intersections of red and blue lines in Fig.\,\ref{fig22}e). It indicates that at these angles the LH and RH modes are degenerate and no handedness-selective optical behavior is expected. When the twist angle $2\alpha$ is not a multiple of $\pi/2$, the symmetry of the whole system is reduced and the degeneracy of the modes is lifted. This leads to emissivity peaks, in which the LH or RH emitter radiates to the far field about 4 orders of magnitude more power than the emitter with the opposite handedness. 

The spatial distribution of the intensity and chirality density of the LH incident wave in the cavity with configurational chirality is shown in Fig.\,\ref{fig23}a,b. It can be seen that, like in Sec.\,\ref{constchirsec} these parameters are almost homogeneous in the entire space between the mirrors. Also, the emissivity of LH chiral emitters is approximately 4 orders of magnitude higher than that of RH emitters almost everywhere in the cavity region (Fig.\,\ref{fig23}c).

Finally, the dispersion of dipoles' emissivity is shown in Fig.\,\ref{fig23}d,e in the vicinity of the $\Gamma$-point. Like in the case of chiral mirrors, as soon as the resonator is designed to a target wavelength and the normal angle, to maintain its properties, deviations of the wavelength and incident/emission angle should be minimized.

\section{Conclusions}
In conclusion, this paper has presented a theoretical study on chiral Fabry-Pérot cavities with mirrors made of van der Waals As$_2$S$_3$ material. By exploiting the anisotropy of As$_2$S$_3$, we have designed cavities with both constitutional and configurational chiralities. In cavities with constitutional chirality, the mirrors themselves possess chirality, resulting in the existence of electromagnetic modes with specific handedness. On the other hand, cavities with configurational chirality exhibit modes of both handednesses, which result from the degeneracy of non-chiral Fabry-Pérot modes. For both types of cavities, we simulated the field distribution of left-handed and right-handed incident waves within the region between the mirrors. At resonant gap sizes, we observed a linearly polarized standing wave with a polarization direction twisted in a helical shape, resulting from the interference between counter-propagating circularly polarized waves of the same handedness. We also simulated the interaction between chiral light, represented by such a twisted standing wave, and matter, modeled by left- and right-handed chiral emitters. These chiral Fabry-Pérot cavities can be adjusted to match the technologically available distance between the mirrors by appropriately tuning their twist angle, making them a promising platform for the interaction of chiral light with chiral matter. The findings of this study contribute to the development of techniques for the selective optical detection of chiral organic molecules, further advancing the field of chiral photonics.

\begin{acknowledgement}

This work was supported by the Russian Science Foundation (Grant No. 22-12-00351). The authors thank M.~V.~Gorkunov and D.~G.~Baranov for fruitful discussion. The authors thank A.S. Slavich and G.A. Ermolaev for providing the optical constants of As$_2$S$_3$. J.-K.~S. was supported by the National Research Foundation of Korea (NRF) grant funded by the Korea government(MSIT) (No. NRF-2022R1A4A1028702),  V.~P.~P. was supported through the NRF Creative Challenge project (No. NRF-2021R1I1A1A01061278).

\end{acknowledgement}

\begin{suppinfo}
\subsection*{Optical constants of As$_2$S$_3$}
\edit{Optical constants of As$_2$S$_3$~\cite{slavich2023exploring} used in our calculations are shown in Fig.\,\ref{as2s3}.}

\begin{figure}[th]
    \centering
    \includegraphics[width=0.7\linewidth]{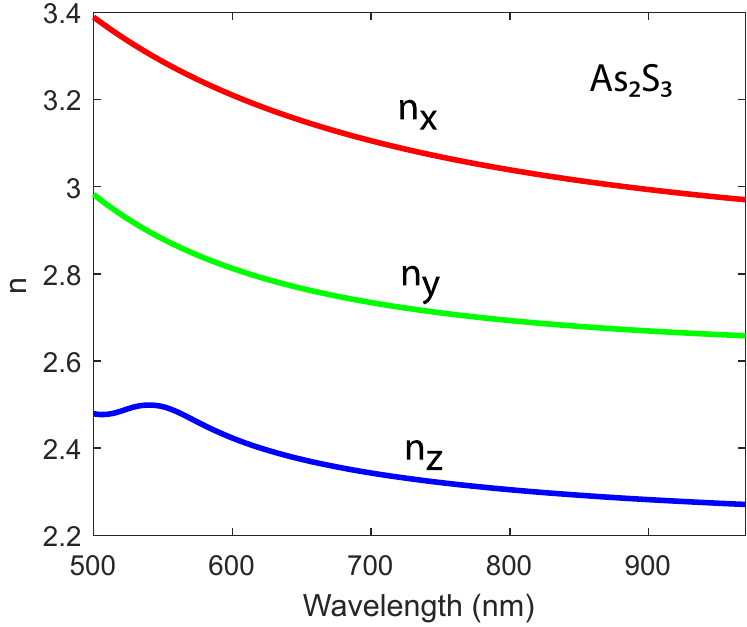}
    \caption{Spectral dependencies of As$_2$S$_3$ refractive indices.}
    \label{as2s3}
\end{figure}

\subsection*{Radiation of a dipole point source}
To calculate the radiation field of oscillating electric and magnetic dipoles, we use the Fourier modal method in the scattering matrix form~\cite{Tikhodeev2002b} (also known as Rigorous Coupled Wave Analysis~\cite{moharam1995formulation}). The power flow of dipoles' radiation through the horizontal planes bounding the resonator at opposite sides is calculated as a $z$-projection of the Poynting vector integrated over the structure period. At a given radiation direction, integration in real space can be replaced with summation over the Floquet-Fourier harmonics:
\begin{equation}
    P = \frac{c}{16\pi}(\mathrm{E}^\dagger_x \mathrm{H}_y + \mathrm{E}_x \mathrm{H}^\dagger_y - \mathrm{E}^\dagger_y \mathrm{H}_x - \mathrm{E}_y \mathrm{H}^\dagger_x),
\end{equation}
where $\mathrm{E}_{x,y}$ and $\mathrm{H}_{x,y}$ are the hypervectors of Floquet-Fourier components of electric and magnetic vectors of dipoles' radiation, $c$ is the speed of light, and dagger denotes the conjugate transpose. According to the Fourier modal method formalism, $\mathrm{E}_{x,y}$ and $\mathrm{H}_{x,y}$ are \edit{calculated} from the vector of amplitudes of plane waves propagating in positive and negative $z$-directions:
\begin{equation}
    \begin{bmatrix}
    \mathrm{E}_x\\\mathrm{E}_y\\\mathrm{H}_x\\\mathrm{H}_y
    \end{bmatrix}_\mathrm{u}=
    \mathbb{F}_\mathrm{u}
    \begin{bmatrix}
    \vec{\mathrm{o}} \\ \vec{\mathrm{u}}
    \end{bmatrix}\hspace{30pt}
    \begin{bmatrix}
    \mathrm{E}_x\\\mathrm{E}_y\\\mathrm{H}_x\\\mathrm{H}_y
    \end{bmatrix}_\mathrm{d}=
    \mathbb{F}_\mathrm{d}
    \begin{bmatrix}
    \vec{\mathrm{d}} \\ \vec{\mathrm{o}}
    \end{bmatrix},
\end{equation}
where $\mathbb{F}_\mathrm{u,d}$ is a material matrix of the layer where the radiation field is calculated; indices \textit{u} and \textit{d} mean that corresponding quantities are taken in the upper or lower semi-infinite media; $\vec{\mathrm{u}}$ and $\vec{\mathrm{d}}$ are the outgoing vectors of amplitudes \edit{and $\vec{\mathrm{o}}$ is the zero vector of the same size as $\vec{\mathrm{u}}$ and $\vec{\mathrm{d}}$. Amplitudes $\vec{\mathrm{u}}$ and $\vec{\mathrm{d}}$ can be found from the hypervectors of oscillating currents induced by oscillating dipole moments that cause additional internal boundary conditions \cite{Graham2000, lindell1994electromagnetic}.} The expression for the vectors of amplitudes reads ~\cite{whittaker1999scattering, whittaker2000inhibited, taniyama2008s, lobanov2012emission, fradkin2019fourier, fradkin2020nanoparticle, dyakov2020vertical}:
\begin{align}
\begin{split}
 \vec{\mathrm{u}} &= \mathbb{S}^{\mathrm{u}}_{22}\left(\mathbb{S}^{\mathrm{d}}_{21}\mathbb{S}^{\mathrm{u}}_{12}-\mathbb{I}\right)^{-1}\left(\vec{\mathrm{A}}_\mathrm{d}-\mathbb{S}^{\mathrm{d}}_{21}\vec{\mathrm{A}}_\mathrm{u}\right),\\      \vec{\mathrm{d}} &= \mathbb{S}^{\mathrm{d}}_{11}\left(\mathbb{I}-\mathbb{S}^{\mathrm{u}}_{12}\mathbb{S}^{\mathrm{d}}_{21}\right)^{-1}\left(\vec{\mathrm{A}}_\mathrm{u}-\mathbb{S}^{\mathrm{u}}_{12}\vec{\mathrm{A}}_\mathrm{d}\right).
 \label{eqn:updn2}
\end{split}
\end{align}
In formula \eqref{eqn:updn2} $\mathbb{S}^{\mathrm{u,d}}$ are the upper and lower partial scattering matrices \cite{Tikhodeev2002b, lobanov2012emission} calculated at a given frequency and emission angle; $\vec{\mathrm{A}}_\mathrm{u}$ and $\vec{\mathrm{A}}_\mathrm{d}$ are the hypervectors of oscillating current density that are found from the Floquet-Fourier components of electric and magnetic dipole moments by the use of the material matrix $\mathbb{F}$ of the layer where the dipole is located:
\begin{equation}
    \label{jujd}
    \begin{bmatrix}
    \vec{\mathrm{A}}_\mathrm{u} \\ \vec{\mathrm{A}}_\mathrm{d}
    \end{bmatrix}
    =\mathbb{F}^{-1} \begin{bmatrix}
    +K_x\mathrm{P}_z+k_0\mathrm{M}_y\\
    +K_y\mathrm{P}_z-k_0\mathrm{M}_x\\
    +k_0\mathrm{P}_y-K_x\mathrm{M}_z\\
    -k_0\mathrm{P}_x-K_y\mathrm{M}_z\\
    \end{bmatrix}\frac{1}{k_0}.
\end{equation}
Here $K_{\mathrm{x,y}}$ are the diagonal matrices of $x$- and $y$-components of photon in-plane wavevector vector of different diffraction orders; $k_0$ is absolute value of the photon wavevector in vacuum; $\tilde\varepsilon$ is a 3$\times$3 block matrix with components that evolve from the Fourier transform of dielectric permittivity tensor \cite{Weiss2009a} calculated in accordance with Li's factorization rules \cite{li1997new}. Note that in formula \eqref{jujd}, matrix elements $\tilde\varepsilon_{13}$, $\tilde\varepsilon_{23}$, $\tilde\varepsilon_{31}$, $\tilde\varepsilon_{32}$ are assumed to be zero. The corresponding hypervectors of electric and magnetic dipole moments positioned at a coordinate $\vec{r}_0$ have the form:
\begin{align}
    \label{peq}
    \begin{split}
    \mathrm{P}_\alpha&=p_\alpha e^{-i\vec{r}_0\left(\vec{k}_\parallel+\vec{G}_{\gamma}\right)},\\
    \mathrm{M}_\alpha&=m_\alpha e^{-i\vec{r}_0\left(\vec{k}_\parallel+\vec{G}_{\gamma}\right)},
    \end{split}
\end{align}
where $p_\alpha$ and $m_\alpha$ are the components of electric and magnetic dipole moments in real space, $\vec{k}_\parallel=\left[k_x,k_y\right]$ is the in-plane wavevector, $\vec{G}_{\gamma}$ is a vector in reciprocal space representing $\gamma$-th harmonic. In Figs.\,\ref{fig12}--\ref{fig23}, we average over the dipoles' orientation, considering electric dipole moments  $\vec{p}_0=\left[1, 0, 0\right]$ and $\vec{p}_0=\left[0, 1, 0\right]$ and the corresponding magnetic dipole moments in accordance with formula \eqref{pmdipoles}. 

\end{suppinfo}

\providecommand{\latin}[1]{#1}
\makeatletter
\providecommand{\doi}
  {\begingroup\let\do\@makeother\dospecials
  \catcode`\{=1 \catcode`\}=2 \doi@aux}
\providecommand{\doi@aux}[1]{\endgroup\texttt{#1}}
\makeatother
\providecommand*\mcitethebibliography{\thebibliography}
\csname @ifundefined\endcsname{endmcitethebibliography}  {\let\endmcitethebibliography\endthebibliography}{}

\end{document}